\documentclass[final,5p,twocolumn]{elsarticle}
\pdfoutput=1

\usepackage[utf8]{inputenc}

\usepackage{lineno}
\usepackage{graphicx}
\usepackage[caption=false,font=footnotesize]{subfig}

\usepackage{url}

\begin{document}


\begin{frontmatter}

\title{Locating the avalanche structure and the origin of breakdown generating
charge carriers in silicon photomultipliers by using the bias dependent breakdown probability}

\author[gt]{Adam~Nepomuk~Otte\corref{mycorrespondingauthor}}
\ead{otte@gatech.edu}
\author[gt]{Thanh~Nguyen}
\author[gt]{Joel~Stansbury}
\address[gt]{School of Physics \& Center for Relativistic Astrophysics, Georgia
Institute of Technology,\\
837 State Street NW, Atlanta, GA 30332-0430, USA.
}

\cortext[mycorrespondingauthor]{Corresponding author}

\begin{abstract}
We present characterization results of two silicon photomultipliers; the Hamamatsu
LVR-6050-CN and the Ketek PM3325 WB. With our measurements of the bias
dependence of the breakdown probability we are able to draw conclusions about
the location and spatial extension of the avalanche region. 
For the KETEK SiPM
we find that the avalanche region is located close to the surface. In the
Hamamatsu SiPM the high-field region is located $0.5\,\mu$m below the
surface, while the volume above is depleted almost until the surface.
Furthermore, for the Hamamatsu SiPM we find that charge carriers
produced by optical-crosstalk photons enter a cell below the avalanche
region as opposed to an earlier device where most of the photoelectrons enter a cell
from above. 
The present paper is an attempt to spur further interest in the use of the bias
dependence of the breakdown probability and establish it as a standard tool not
only to determine the location of the high-field region but also to determine
the origin of charge carriers relative to the high-field region. With the
knowledge of where the charges come from it should be possible to further improve
the optical crosstalk, dark count, and afterpulsing characteristics of SiPM.
\end{abstract}


\end{frontmatter}

\section{Introduction}
The silicon photomultiplier (SiPM) has evolved into an
established photodetector technology. They are used in high-energy physics
\cite{Ogawa2017,Moreau2010,Mannel2013}, astroparticle physics
\cite{Anderhub2013,Otte2015}, medical imaging \cite{Otte2005,Spanoudaki2007},
and LIDARs \cite{Perenzoni2017,Agishev2013}, to only name a few areas of
application. One key factor to the success of SiPMs is the continuing effort
made by manufacturers to reduce nuisance parameters like dark-count rate,
afterpulsing, and optical crosstalk.

Diagnostic tools are crucial in these efforts as they help to identify means
that further reduce nuisance parameters, which in turn, improves the performance
of SiPMs. One way to diagnose SiPMs is to measure their characteristics as
function of temperature and bias, model the data, and extract physical
meaningful quantities from the model parameters.  We have taken that
approach in earlier work \cite{Otte2016a} and we use it again here. 

In this paper we emphasize the use of the bias dependence of the breakdown
probability, which we already used in \cite{Otte2016a} to determine the origin
of optical crosstalk in a Hamamatsu device and extend it to a discussion of the
location of the avalanche region. The approach is not new, we first presented it
at \cite{Otte2015a} and it was used to characterize FBK devices
\cite{Zappala2016}. Compared to \cite{Zappala2016} we use a parameterization,
which is less dependent on the device specifics as we will discuss in detail.

\section{Devices used in this Study}

The Hamamatsu SiPM is a prototype named LVR2-6050-CN.  The device has an
active area of $6\times6$\,mm$^{2}$ and is composed of $50\,\mu$m sized cells.
For better UV sensitivity the sensor is not covered with a protective layer. The
breakdown voltage at room temperature ($24^\circ$C)  is 38.4\,V and the bias
voltage to achieve a 90\% breakdown probability for 400\,nm photons is about
42\,V (see later). That bias voltage is less than
the 56\,V required for the Hamamatsu LCT5 device we tested in \cite{Otte2016a}.
Whether the lower bias is due to a narrower high-field region in the present
device or due to other changes in the technology we do not know.

The second device is a KETEK PM3325 WB SiPM.\footnote{\url{https://www.ketek.net/store/category/sipm-standard-devices/wb-series/}} It has an active area of
$3\times3$\,mm$^{2}$ and $25\,\mu$m cells. The chip is protected with a
$400\,\mu$m thick glass window. The PM3325 does not feature trenches to suppress
optical crosstalk. The bias voltage to achieve a 90\% breakdown probability
when illuminated with 400\,nm photons is about 32\,V and the breakdown voltage
is 27.5\,V at room temperature.

\begin{figure}[!tb]
  \centering
  \includegraphics*[width=0.8\columnwidth]{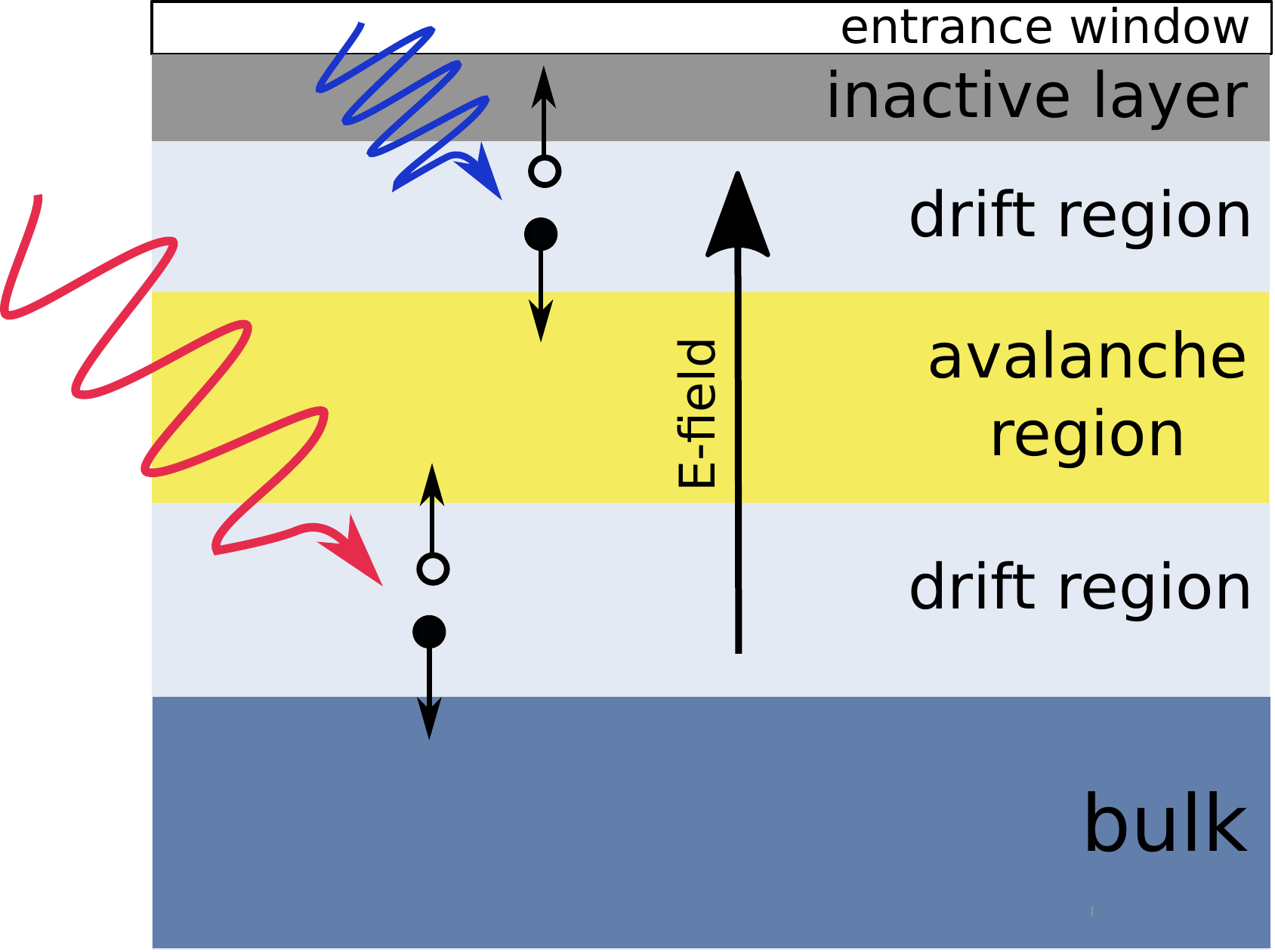}
\caption{
Conceptual cross section of one cell of a \emph{p}-on-\emph{n} SiPM. Blue photons are absorbed mostly before
reaching the avalanche region and an electron (filled circle) drifts down into the
high-field region. Red photons are absorbed mostly after the avalanche region and a
hole (empty circle) drifts up into the high-field region. If the photon is absorbed
in the non-depleted bulk, the hole first has to diffuse into the depleted volume
before it can drift into the avalanche region. 
}
  \label{fig:sketch}
\end{figure}

\section{Probing the Avalanche Structure with Photon Detection Efficiency
Measurements}

The photon detection efficiency (PDE) is one example where the breakdown
probability plays a decisive role.  Depending on the photon's absorption length
and the location and extension of the high-field region, a photon is either
absorbed before the high-field region (blue photons) or after it (red photons).
See Figure \ref{fig:sketch} for a conceptual sketch of one SiPM cell, which
illustrates the situation.

The photon absorption results in the generation of an electron and hole,
which - in case the absorption takes place in the active volume of the cell -
drift in opposite directions due to the electric field in the depleted volume.
If the photon is absorbed after the high-field region in a \emph{p}-on-\emph{n}
structure like the ones studied here, it is the hole that drifts into the high-field region, if the photon
absorbs before the high-field region, it is the electron that drifts down into the avalanche
region.

The probability to initiate a Geiger breakdown is smaller for holes than
for electrons (due to the lower mobility of holes in silicon, e.g.\ \cite{Oldham1972}).
If one could  
measure the probability of a subsequent breakdown as a function of where the
electron/hole pair is released one would, therefore, reverse engineer the location and vertical extension of the
high-field region. 
Such a mapping is indeed possible with bias dependent PDE
measurements as has been shown in \cite{Otte2015a,Zappala2016}.

For the Hamamatsu SiPM we measured the PDE at three wavelengths and for the
KETEK device at four wavelengths.  A description of the setups and procedures
used for the PDE and all other measurements presented here is given in
\cite{Otte2016a}. 

Like in our previous measurements we find that the PDE for
a given wavelength is well fit with the empirical model
\begin{equation}\label{eq:PDE}
PDE(U_{\mbox{\tiny rel}}) = PDE_{\mbox{\footnotesize max}}\left[1-e^{-
\mathcal{O}\cdot U_{\mbox{\tiny
rel}}} \right],
\end{equation}
where $U_{\mbox{\tiny rel}}= \left(U - U_{\mbox{\tiny BD}}\right)/{U_{\mbox{\tiny
BD}}}$ is the relative overvoltage above the breakdown
voltage $U_{\mbox{\tiny BD}}$. $PDE_{\mbox{\footnotesize max}}$ is the PDE in
saturation but is not necessarily the true saturation value because we cannot
measure the PDE at higher bias values. The term in square brackets is the breakdown
probability, which depends only on the product of the relative overvoltage and a 
dimensionless parameter $\mathcal{O}$, which is mostly dependent on whether
an electron or a hole initiates a breakdown as we explain
later.\footnote{$\mathcal{O}$ was jocularly referred to as the \emph{Otte
number} at recent meetings.} 

In that context it is interesting to remark that empirically all the bias
dependent physics of the breakdown is included in one single constant, or in a
linear function when larger relative overvoltages than measured here are taken
into account \cite{Zappala2016}. Because our data is well described with one
constant we do not need to consider the linear function, which would,
furthermore, not be sufficiently constrained by our data. The devices we tested
cannot be operated much beyond the measured voltage range.

While the overall fit function is the same as in \cite{Zappala2016} there are
two differences in its usage. Instead of plotting the breakdown probability as a
function of absolute bias voltage we use the relative overvoltage
$U_{\mbox{\tiny rel}}$. The second difference is that we characterize the
electron/hole initiation probability with $\mathcal{O}$ instead of the voltage
at which the PDE reaches 95\%.  $\mathcal{O}$ and $U_{\mbox{\tiny rel}}$ are inherently less
dependent on the structure of the device and temperature than the absolute bias voltage as we
shall motivate in the following.

The avalanche and breakdown characteristics of a pn-junction are governed by the
ionization rates, which depend strongest on the electric field and much less on
device specifics like the doping profile, doping concentrations, or
temperature \cite{Sze2007}. By parametrizing the breakdown probability as a
function of the average electric field $\bar{E}$ in the high-field region and
not as a function of absolute voltage one arrives at a
parameterization that depends mostly on avalanche physics.
With such a parameterization it should then be possible to extract information
about the breakdown characteristics that can be compared with measurements from
other devices in a meaningful manner.

\begin{figure}[!tb]
   \subfloat[Hamamatsu
LVR2]{\includegraphics*[width=\columnwidth]{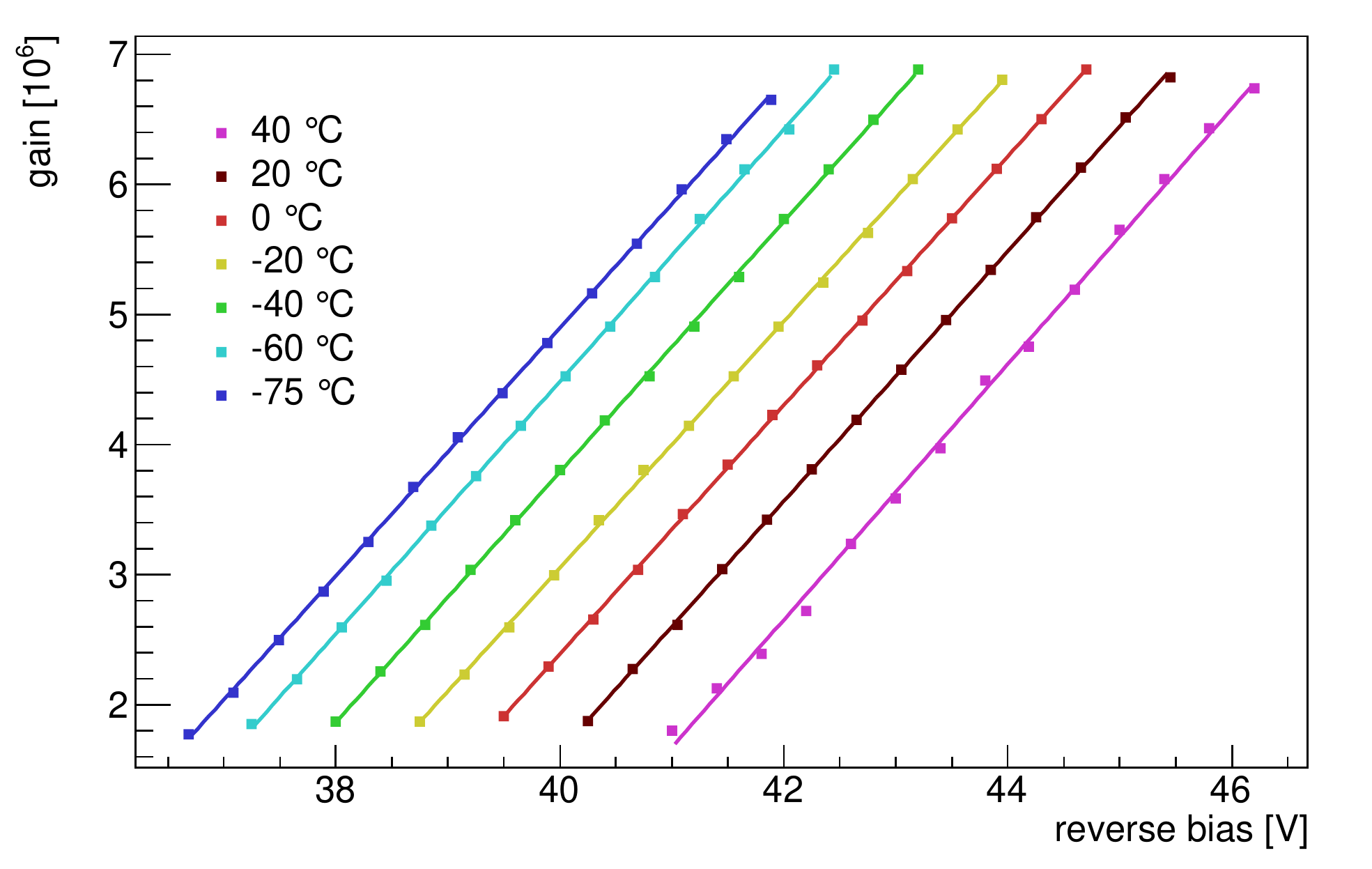}}

   \subfloat[KETEK
PM3325]{\includegraphics*[width=\columnwidth]{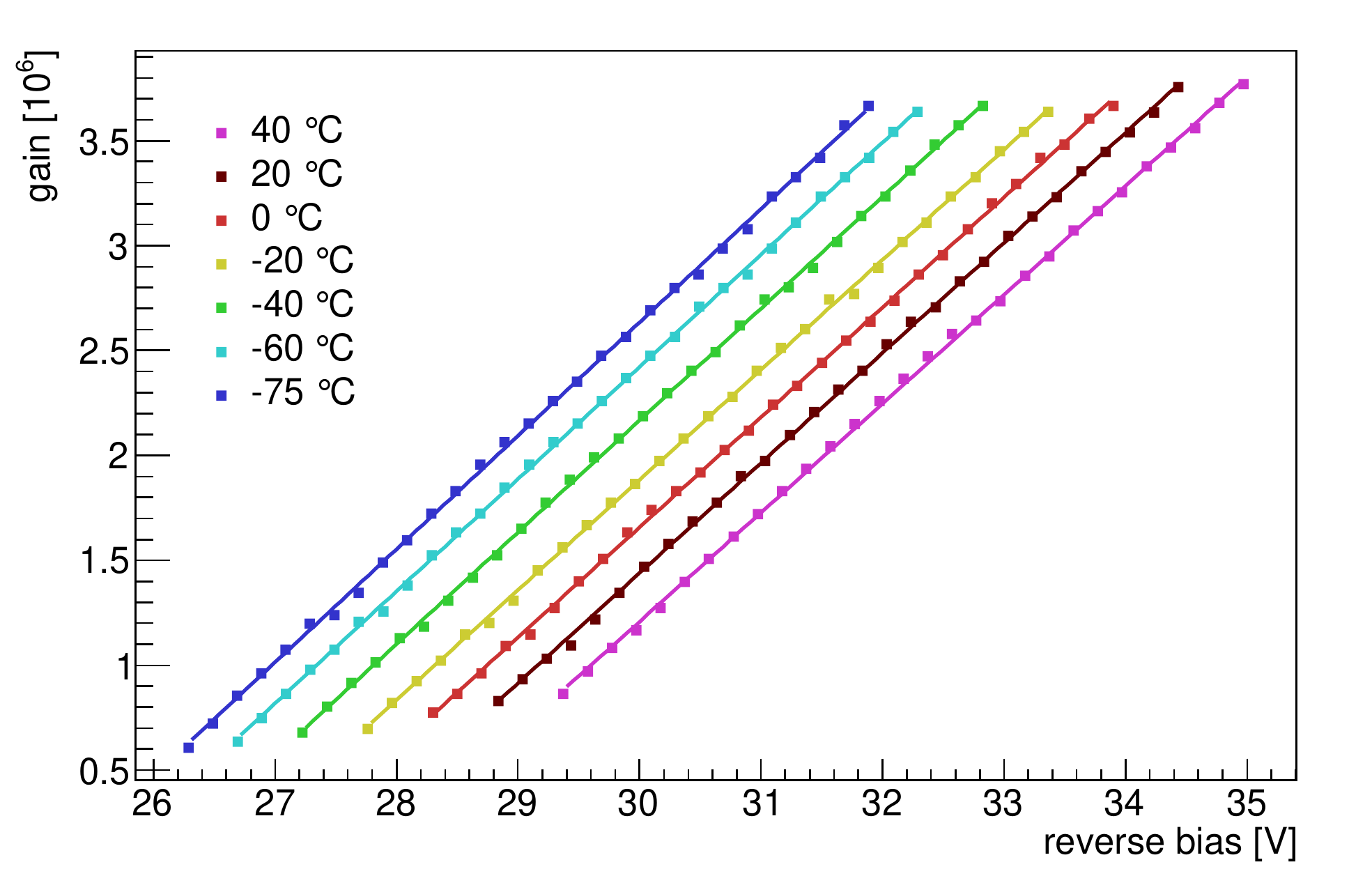}}
\caption{Gain as a function of absolute voltage for seven different
temperatures.
  \label{fig:Gain}
}
\end{figure}

The bias $U = \bar{E}/w$ depends on the device specific
parameter $w$, \emph{i.e.}\ the \emph{effective} width of the high field
region and thus cannot fulfill the task of a device-independent characterization.  The
relative overvoltage $U_{\mbox{\tiny rel}}$, on the other hand, is independent of $w$ and proportional
to $\bar{E}$.
\begin{equation}
 U_{\mbox{\tiny rel}} = \frac{U - U_{\mbox{\tiny BD}}}{ U_{\mbox{\tiny BD}}} =
\frac{\bar{E}\cdot w - \bar{E}_{\mbox{\tiny BD}}\cdot w}{ \bar{E}_{\mbox{\tiny BD}}\cdot w} =
\frac{\bar{E} - \bar{E}_{\mbox{\tiny BD}}}{ \bar{E}_{\mbox{\tiny BD}}}
\end{equation}
where $\bar{E}_{\mbox{\tiny BD}}$ is the electrical field at breakdown averaged
across the high-field region.
We note that $w$ drops out if the width of the depleted region does not change
between breakdown and operating voltage. That assumption holds true for most
available SiPM including the tested devices where the gain as a function of bias
voltage is described by a linear function (see Figure \ref{fig:Gain}).

While, as mentioned above, most of the breakdown characteristics depend on the
electric field, other factors play a role too. $U_{\mbox{\tiny rel}}$ compensates
for some but admittedly, not all of the device and temperature dependencies by
normalizing to $\bar{E}_{\mbox{\tiny BD}}$.  
We can show that at least the temperature dependencies of the breakdown
characteristics are properly taken care of. Optical
crosstalk measurements taken at 100\,K temperature difference fall on top of
each other when plotted as a function of $U_{\mbox{\tiny rel}}$ (see Figure
\ref{fig:PromptOC}), which would not be the case if plotting against $U_{\mbox{\tiny rel}}$ would
not compensate for temperature dependencies. The picture is very different when
optical crosstalk is plotted as a function of absolute voltage.

Using $U_{\mbox{\tiny rel}}$ in the argument of the exponential function of the
breakdown probability can be viewed as a Taylor series expansion about the
critical electric field $\bar{E}_{\mbox{\tiny BD}}$. The linear coefficient in
the expansion is $\mathcal{O}$, the constant term is obviously zero or so small
that it is not relevant, higher order terms can be relevant \cite{Zappala2016}. $\mathcal{O}$ thus parameterizes the electric field
dependence of the breakdown, which as we have discussed above does not depend
much on the device specifics. $\mathcal{O}$ can thus be compared between
devices, contrary to the absolute voltage when the breakdown probability reaches
95\%.

\begin{figure*}[!tb]
  \centering
  \subfloat[Hamamatsu LVR2]{\includegraphics*[width=\columnwidth]{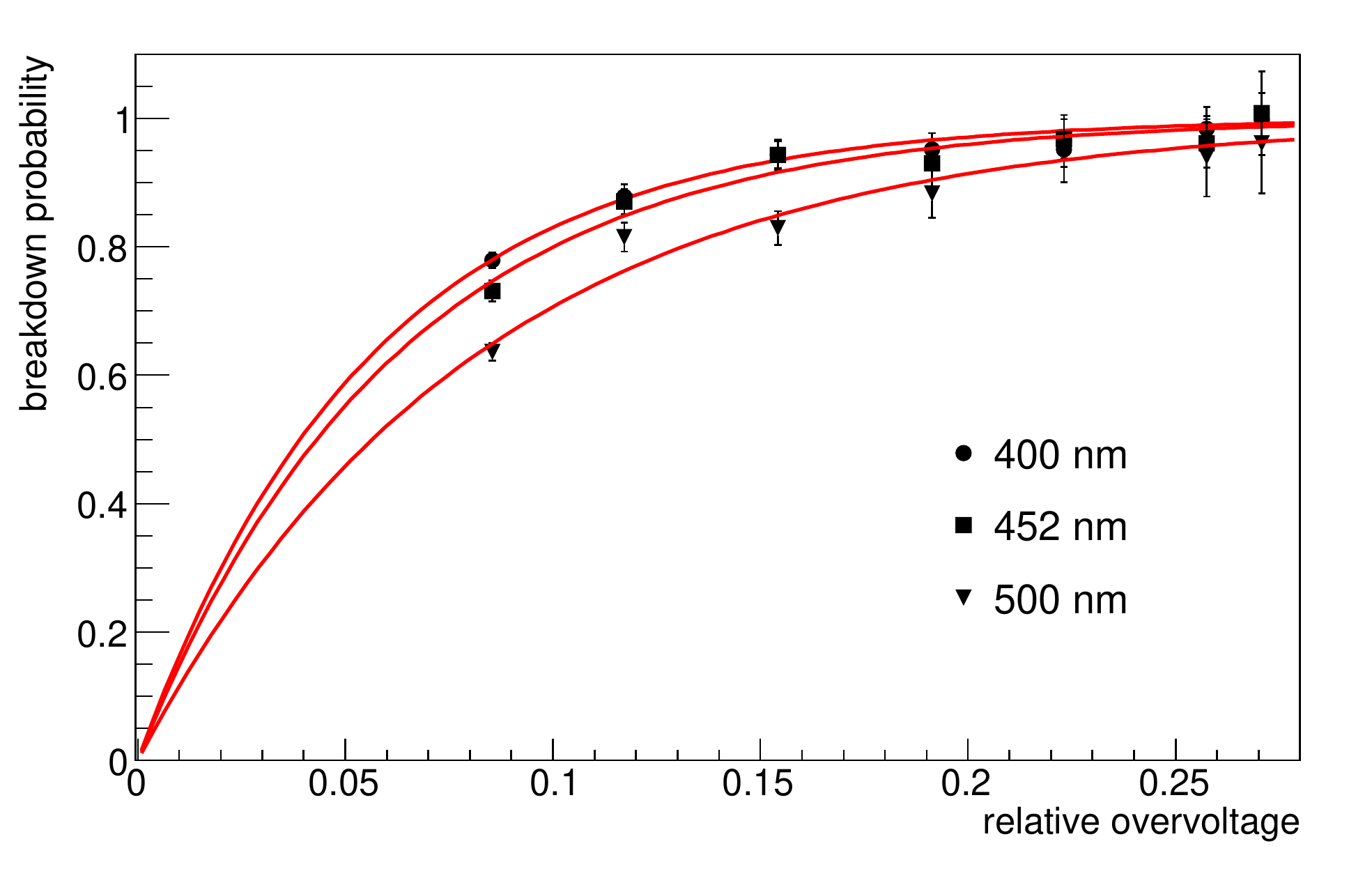}}
  \subfloat[KETEK PM3325]{\includegraphics*[width=\columnwidth]{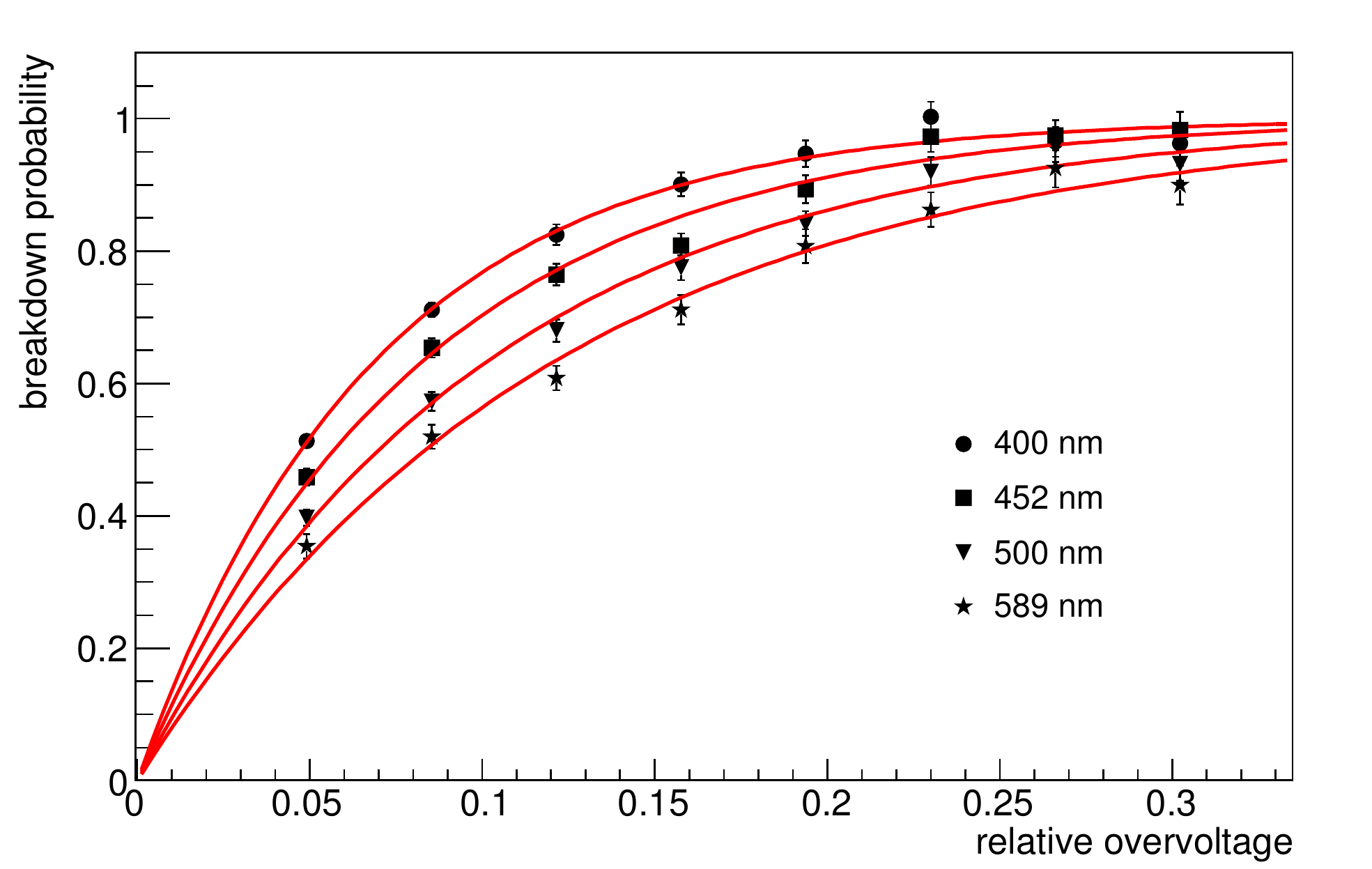}}
\caption{Breakdown probability versus relative overvoltage for the two tested
devices. The lines are fits to the data points with the model described
in the text.
}
  \label{fig:BD}
\end{figure*}

 But the breakdown probability depends strongly on
whether an avalanche is initiated by electrons or holes and it is, therefore,
expected that $\mathcal{O}$ changes with changing electron/hole breakdown
initiation ratio.
Figure \ref{fig:BD} shows the breakdown probability derived from the PDE
measurements, i.e.\ the PDE divided by $PDE_{\mbox{\footnotesize max}}$.
The solid lines depict the best fit parameterizations of the breakdown
probability, which all yield fit probabilities of 30\% or better. 
The fitted values of $\mathcal{O}$ are listed in Table
\ref{tab:ottenumb} together with the corresponding photon absorption lengths.
The 589\,nm light source was not available for the measurement of the
Hamamatsu device.

The value of $\mathcal{O}$  decreases with increasing photon wavelength for each
device, which is a testimony to the fact that the breakdown probability shifts
from majority electron to majority hole initiated breakdowns. 
$\mathcal{O}$ thus shows a clear dependence on the ratio of electron to hole
initiated breakdowns. 

\begin{table}
\caption{Values of $\mathcal{O}$ derived from PDE measurements at different
wavelength for the two devices. The second column gives the absorption length of
photons with the wavelength given in the first column.\label{tab:ottenumb}}
\centering
\begin{tabular}{c|c|c|c}
\hline
Wavelength & Absorp. length&\multicolumn{2}{c}{
$\mathcal{O}$}\\\cline{3-4}
$[$nm$]$ & $[\mu$m$]$&Hamamatsu & KETEK \\\hline
400 & 0.082 &$17.7\pm0.6$  & $14.6\pm0.3$  \\\hline
452 & 0.43 & $16.1\pm0.6$ & $12.1\pm0.3$ \\\hline
500 & 0.91 & $12.3\pm0.4$ & $9.9\pm0.2$ \\\hline
589 & 2.0 & N/A & $8.3\pm0.2$ \\\hline
\end{tabular}
\end{table}

The absolute value of $\mathcal{O}$ should also depend on the dimensions of the 
avalanche region, which we do not know and thus cannot explore further.
For the time being, we resort to the assumption that the dependence
of $\mathcal{O}$ on the width of the avalanche region is small compared to the
observed change with photon-wavelength and can be neglected. 
How valid that assumption is needs to be shown in the future on devices
with known dimensions of the high field region.
The avalanche regions of the two tested devices probably have fairly similar widths, which we infer from
the similarities of their respective breakdown voltages, which are 26.8\,V and
37.5\,V at 0\,$^{\circ}$C for the KETEK and Hamamatsu SiPM, respectively.

For the Hamamatsu device $\mathcal{O}$ is 12 for photon
absorption lengths of $0.9\,\mu$m while the KETEK SiPM yields the same number for
absorption lengths of $0.4\,\mu$m. If the difference in absorption lengths is
taken at face value and $\mathcal{O}$ does not depend strongly on details of the
two structures, it follows that the avalanche region is located $0.5\,\mu$m
deeper in the Hamamatsu SiPM than in the KETEK SiPM. 

Two more observations are that a) in between absorption lengths $0.08\,\mu$m and
$0.4\,\mu$m, $\mathcal{O}$ changes little in the Hamamatsu SiPM, while it
changes much more in the KETEK SiPM. And b) $\mathcal{O}$ never reaches as high
a value in the KETEK SiPM as in the Hamamatsu SiPM.  Under the assumption that
$\mathcal{O}$ does not depend strongly on details of the two structures, we
interpret both observations as evidence for a location of the avalanche region
in the KETEK SiPM that is right below the surface and that already for 400\,nm
photons a significant fraction of photons absorb after the avalanche region.  In
the Hamamatsu SiPM, on the other hand, the passive region right below the
surface and before the drift volume starts is thinner than in the KETEK device.
Thus more photons are absorbed and mostly electrons drift into the high-field
region also for $<400$\,nm photons.

\begin{figure}[!tb]
  \centering
  \includegraphics*[width=\columnwidth]{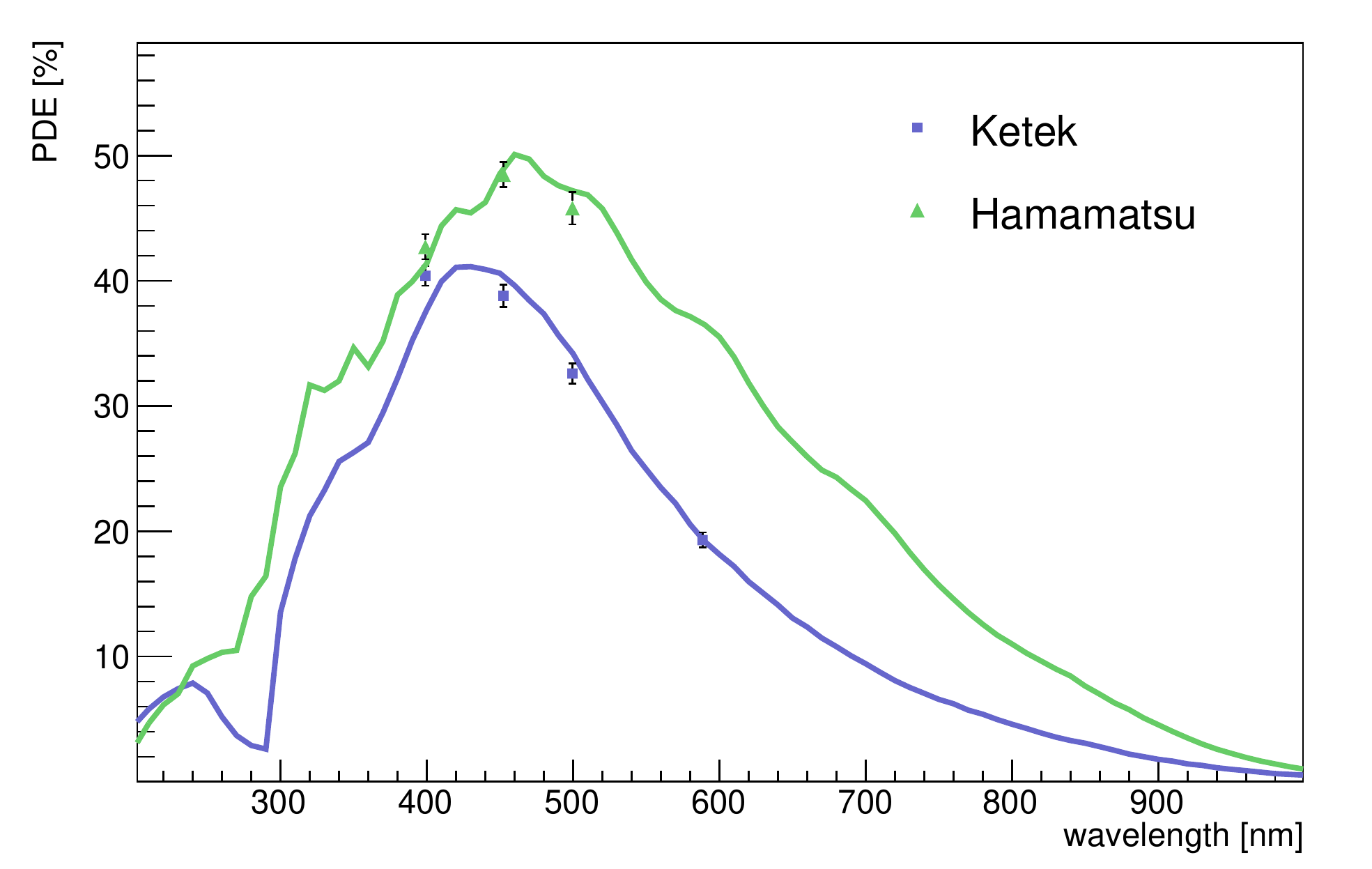}
\caption{
PDE vs.\ wavelengths of the two SiPMs from 200\,nm to 1000\,nm. 
For the measurement, the bias voltage for each devices is chosen such that the
breakdown probability for 400\,nm photons is 90\%. The spectral response measurement
is fit to the PDE measurements denoted by the data points.
}
  \label{fig:PDE}
\end{figure}

In that scenario it is expected that the spectral response of the Hamamatsu
device is higher below 400\,nm because of the larger active volume above the
high field region and thinner passive area. It is also expected that the response of the KETEK SiPM peaks at lower
wavelengths than in the Hamamatsu SiPM because the breakdowns change to hole
dominate ones for shorter wavelengths in the KETEK device than in the Hamamatsu
one. That is indeed what we
observe. Figure \ref{fig:PDE} shows the spectral response of the two devices
measured with the setup explained in \cite{Otte2016a}.

\section{Where Optical Crosstalk Photons enter a Cell}

In this section we discuss how $\mathcal{O}$ can be used to determine
where optical crosstalk photons enter a cell.  Optical crosstalk (OC) is caused by photons
that are emitted in the breakdown of one cell and propagate into a neighboring
cell where they initiate an additional breakdown.  One distinguishes two types
of OC (see e.g.\ \cite{Buzhan2009} ) In case the photon absorbs in the active (depleted) volume of a cell,
the additional breakdown happens nearly simultaneous to the first breakdown, which is why that type of OC is called prompt or direct OC. If
the photon is absorbed in a non-depleted region, e.g.\ in the bulk, the
generated charges first have to diffuse into the depleted volume before they can
initiate a breakdown . The diffusion time $\Delta t$ can take several tens of
nanoseconds depending on the distance $d$ between the location of the photon
absorption and the border to the active volume of the cell; $\Delta t \propto
\sqrt{d}$. But it can also be just a fraction of a nanosecond if the photon
absorbs close to the border.

How well the two types of OC can be separated depends on how well two subsequent
pulses can be separated in the measurement.  Any prompt OC measurement is thus
always a combination of \emph{true} prompt OC events and delayed OC events that
have a time delay, which is below the capability of the measurement setup to
resolve two overlapping pulses. Two pulses can be identified as
such in our setup, if they are more than two nanoseconds apart. 

\begin{figure*}[!tb]
  \centering
   \subfloat[Hamamatsu
LVR2]{\includegraphics*[width=\columnwidth]{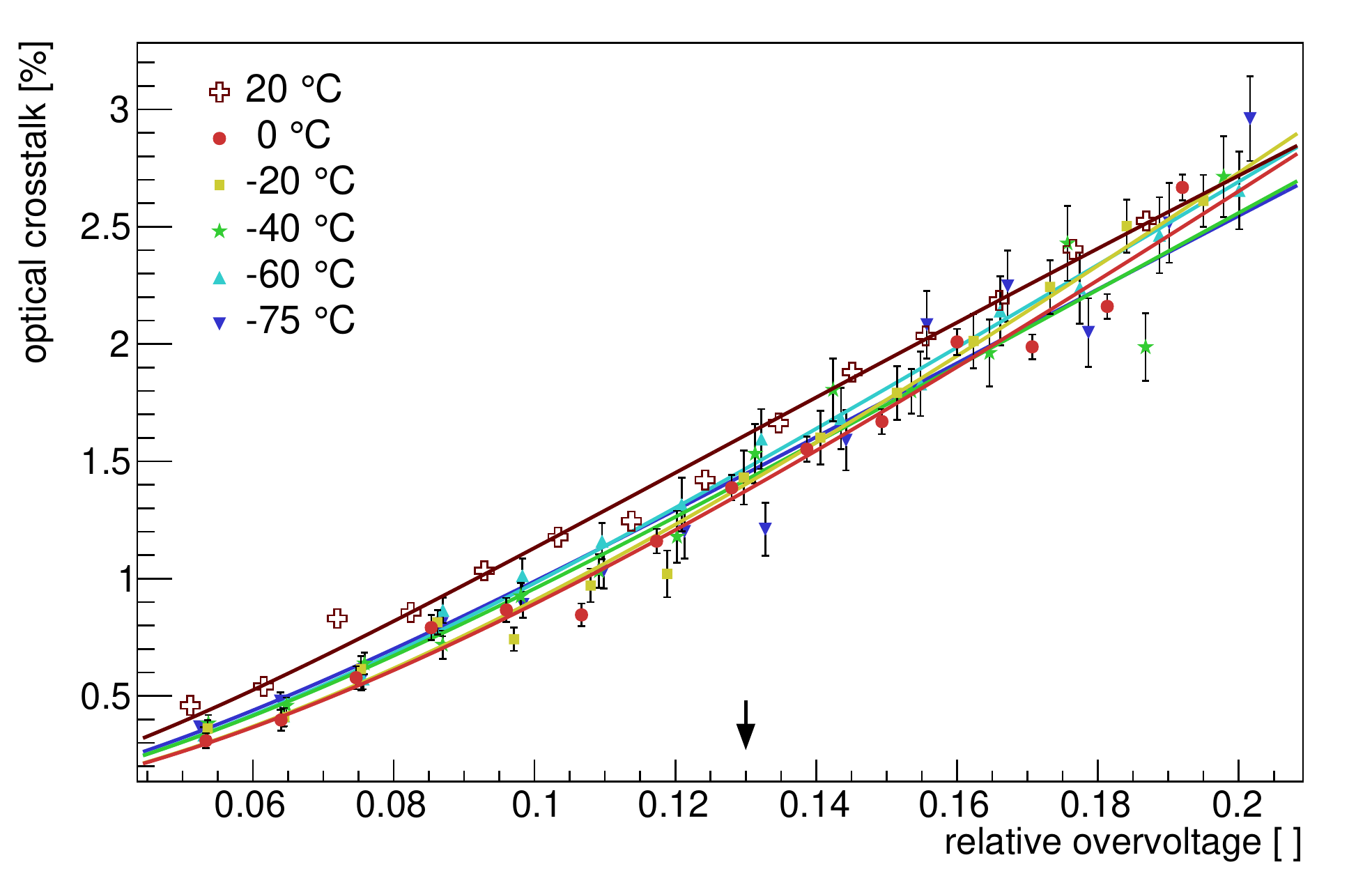}}
   \subfloat[KETEK PM3325]{\includegraphics*[width=\columnwidth]{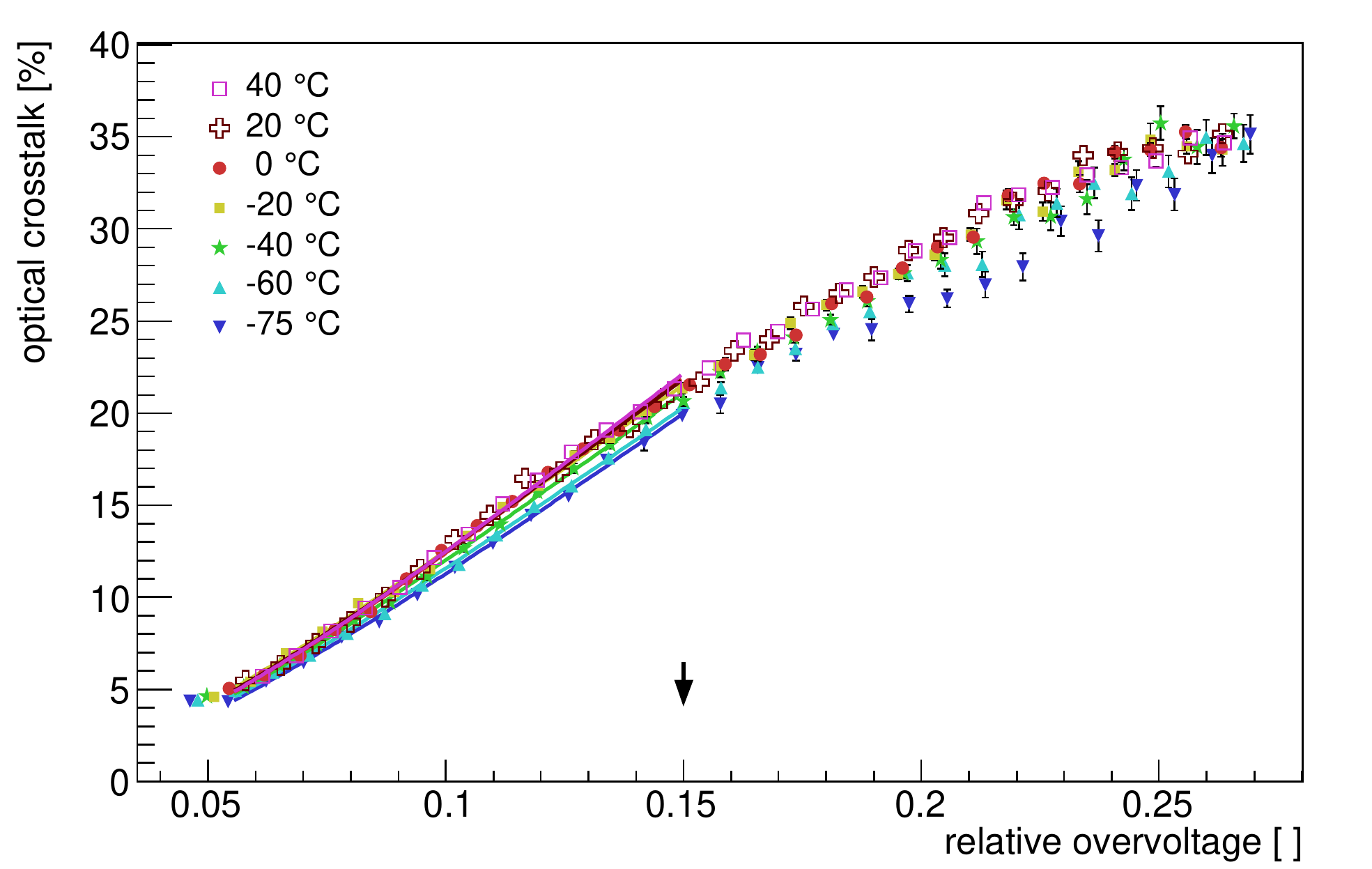}}
\caption{Prompt optical crosstalk of the two tested SiPMs. The black arrow marks
the relative overvoltage at which both devices yield a 90\% breakdown
probability for 400\,nm photons. 
  \label{fig:PromptOC}
}
\end{figure*}

Figure \ref{fig:PromptOC} shows the prompt OC of the two devices recorded at
seven temperatures between $-75^{\circ}$C and $40^{\circ}$C. In this and
subsequent measurements, OC is quantified as the probability that the breakdown
of one SiPM cell causes one or more other cells to break down too. For the Hamamatsu
device we discarded the measurement at $40^{\circ}$C because the contamination
from pile-up of uncorrelated dark counts was too large and could not be reliably
subtracted. For all other measurements, the accidental pile-up within a 2\,ns
time window could be subtracted by assuming that the number of dark counts in a
given time interval are Poisson distributed. After the correction, all OC curves of one
device fall on top of each other, as expected.

We now compare the OC of the two devices at the bias where the breakdown
probability for 400\,nm photons is 90\%.\footnote{We do not imply that this
operating point is optimal for an application but it allows for an unbiased
comparison.} The arrow in each panel marks the corresponding relative
overvoltage. The KETEK device has a fairly high optical crosstalk of $\sim20$\%,
which is not surprising because it does not have trenches to prevent photons
from propagating into neighboring cells.  The prompt OC in the Hamamatsu device,
on the other hand, is only 1.5\%, which is an impressive improvement compared to
past developments \cite{Otte2016a}. 

In \cite{Otte2016a} we showed that a valid model of the  optical
crosstalk probability vs.\ relative overvoltage is 
\begin{equation}\label{OCf}
OC(U_{\mbox{\footnotesize rel}}) = f\cdot  C_{\mbox{\footnotesize eff}}\cdot U_{\mbox{\footnotesize rel}}\cdot U_{\mbox{\footnotesize BD}}\cdot \gamma\cdot
\left[1-e^{\left(-\mathcal{O}\cdot U_{\mbox{\footnotesize rel}}\right)}\right]\,.
\end{equation} where we use $f=3\cdot10^{-5}$ from \cite{NepomukOtte2009a} as the number of photons produced per
charge carrier in the avalanche that can also cause OC. We note that other measurements of the photon
intensity exist, e.g.\ \cite{Lacaita1993,Mirzoyan2009}, but those also include
spectral components, which are irrelevant for OC, either because the photon
absorption lengths are too long (photons do not absorb in the device) or too
short (photons absorb in the same cell they are emitted from). $C_{\mbox{\footnotesize eff}}\cdot
U_{\mbox{\footnotesize rel}}\cdot U_{\mbox{\footnotesize BD}}$ is the gain of
the SiPM, and $\gamma$ is a figure of merit that quantifies what fraction of the
photons produced in a breakdown make it into a neighboring cell. The term in
square brackets is the breakdown probability already discussed in the previous
section. 

The OC data in Figure \ref{fig:PromptOC} are fit with that model. For the fit we
fixed the cell capacitance $C_{\mbox{\footnotesize eff}}$ at 84\,fF and 154\,fF
and the breakdown voltage at 26.8\,V and 37.5\,V at $0^\circ$C, for the KETEK
and Hamamatsu SiPM, respectively.  The capacitance and breakdown voltages had
been measured as described in \cite{Otte2016a}. The breakdown voltage is found to
increase by about 0.1\%/$^\circ$C in both devices.

The Hamamatsu OC measurements can be fit over the entire measured range with an
acceptable fit probability. For the KETEK device, we had to restrict the upper
end of the fit range to a relative overvoltage of 0.15, i.e.\ OC of less 
than 20\%, in order for the fit to yield an acceptable fit probability. It is evident from
the KETEK data points, that the OC data turn over in what seems to be a saturating
behavior. An explanation for this behavior is that for large OC
of more than 20\% and the cell size of the device, the probability of more than one
OC photon being absorbed in the same
cell cannot be neglected anymore. That effect is not included in the fit model.

\begin{table}
\caption{Best Fit Values for $\gamma$ Obtained From Fitting the Prompt Optical Crosstalk Measurements Shown in Fig.\  \ref{fig:PromptOC}.
Also shown is the $\mathcal{O}$ value for each fit. The last three rows give the
values obtained from \cite{Otte2016a}.}
\centering
\begin{tabular}[!htb]{c|c|c|c}
Device&Temp.&$\gamma$&$\mathcal{O}$\\\hline\hline
Hamamatsu&-75$^{\circ}$C&0.012$\pm$0.001&13.9$\pm$1.3\\
 LVR2  &-60$^{\circ}$C&0.014$\pm$0.001&9.9$\pm$1.4\\
   &-40$^{\circ}$C&0.013$\pm$0.001&10.9$\pm$1.6\\
   &-20$^{\circ}$C&0.018$\pm$0.003&6.8$\pm$1.2\\
   &0$^{\circ}$C&0.017$\pm$0.001&7.2$\pm$0.9\\
   &20$^{\circ}$C&0.014$\pm$0.001&16$\pm$0.8\\\hline
KETEK &-75$^{\circ}$C&0.347$\pm$0.009&12.9$\pm$0.7\\
PM3325 WB   &-60$^{\circ}$C&0.355$\pm$0.008&13.7$\pm$0.8\\
   &-40$^{\circ}$C&0.378$\pm$0.009&13.5$\pm$0.7\\
   &-20$^{\circ}$C&0.384$\pm$0.007&15.3$\pm$0.7\\
   &0$^{\circ}$C&0.42$\pm$0.01&12.9$\pm$0.7\\
   &20$^{\circ}$C&0.415$\pm$0.008&13.8$\pm$0.6\\
   &40$^{\circ}$C&0.442$\pm$0.007&12.8$\pm$0.4\\\hline
Hamamatsu LCT5&&0.077$\pm$0.001&13$\pm$0.2\\
SensL J-Series&&0.126$\pm$0.002&8.5$\pm$0.1\\
FBK NUV-HD&&0.557$\pm$0.002& N/A\\
\end{tabular}
\label{OCvals}
\end{table}

Table \ref{OCvals} lists the values for $\gamma$ from the fits. The
average values from our previously measured devices are also listed
\cite{Otte2016a}. Comparing the numbers it is evident that the structure of the
LVR2 device is 5.5 times better than the LCT5 device in preventing photons from
crossing cells.  The value for $\gamma$ is 0.014, i.e.\ 1.4\% of all
photons make it into a neighboring cell where they can cause
optical crosstalk. In the KETEK SiPM, between 35\% and 44\% of the photons cause optical crosstalk. 

The second factor that determines the amount of OC is the product of breakdown
voltage and cell capacitance, which is 2.25\,pF$\cdot$V for the KETEK and
5.78\,pF$\cdot$V for the Hamamatsu SiPM. It is a figure of merit that is
proportional to the charge generated in an avalanche. Minimizing the figure of
merit by designing
devices with small breakdown voltage and/or small cell capacitance
minimizes OC while retaining good breakdown characteristics, which are
governed by $U_{\mbox{\tiny rel}}$.  

This time it is the KETEK
SiPM that outperforms the Hamamatsu device by a factor of 2.6 because of its smaller cell capacitance. However,
the Hamamatsu SiPM has a two times smaller cell capacitance per cell area. We
would thus expect that the product of cell capacitance and breakdown voltage for
an LVR2 with $25\,\mu$m cells will be two times lower than for the KETEK device.
This assumes that the cell capacitance scales linear with area, which is not
necessarily the case as edge effects become important for small cell sizes.

The fit results also allow us to draw conclusions about the location where the
crosstalk producing photons are absorbed relative to the avalanche region. For the
previously tested Hamamatsu LCT5 SiPM we could show that the majority of these
photons are absorbed above the avalanche region \cite{Otte2016a}. The $\mathcal{O}$
value we obtained then was $\sim26$.  That interpretation
was confirmed by Hamamatsu, who found that these photons exit the silicon and
reflect off the boundary between the protective layer and the ambient air back into a
cell. 

In the  Hamamatsu SiPM studied here that contribution to the prompt OC has been
successfully suppressed by eliminating the protective epoxy layer. The same
conclusion comes from the interpretation of $\mathcal{O}$. The best fit
value for $\mathcal{O}$ is about $10\pm1$ in all fits of the optical crosstalk
but the one for $20^\circ$C, where the fit probability is $10^{-7}$ due to a
contamination from random dark counts and can thus be
safely ignored (see Table \ref{OCvals}). 
The average value can be compared with the ones we found 
from the different PDE measurements (Table \ref{tab:ottenumb}). A small
$\mathcal{O}$ value like 10 corresponds to heavily hole initiated breakdowns, which
means that the OC photons must be absorbed below the high-field region. According
to Table \ref{tab:ottenumb} that is the case if the OC photons absorb in a depth
$>1\,\mu$m below the surface.

Three scenarios come to mind that can explain how optical crosstalk photons
can be absorbed at such depths. The first scenario is that some photons manage to
penetrate the trench between cells. That scenario is unlikely because
photons would absorb uniformly across the cell, i.e.\ absorb above and
below the avalanche region and, in consequence, result in 
values for $\mathcal{O}$ larger than 10 because the occurring breakdowns would
be electron and hole initiated.  The second scenario is that some photons with
long absorption lengths still bounce off the air-SiPM interface and are absorbed deep
inside the device, i.e.\ mostly below the avalanche structure. The third, and
our preferred scenario is that photons cross into a neighboring cell below the
trench and are absorbed below the avalanche structure.

In the second and third scenario photons can be absorbed in the bulk and the
generated holes diffuse into the active volume where they cause delayed OC (see
next section). If
the diffusion time is less than 2\,ns and thus below the resolving time of our
setup, the delayed OC would be misidentified as a prompt OC event. If the
photons are absorbed in the active volume below the avalanche region a prompt OC would
be caused.

The fit result for the KETEK SiPM yields an $\mathcal{O}$ number of
$13.6\pm0.7$. Comparing that value with the $\mathcal{O}$  numbers in Table
\ref{tab:ottenumb} lets us conclude that the majority of the photons absorb
equally distributed across the avalanche region and thus produce an equal amount of
electron and hole dominated breakdowns. That result is not surprising as the
device does not have trenches in between cells, which would prevent photons to
travel directly from the avalanche region where they are produced into a
neighboring one.

\begin{figure*}[!tb]
  \centering
   \subfloat[Hamamatsu
LVR2]{
  \includegraphics*[width=\columnwidth]{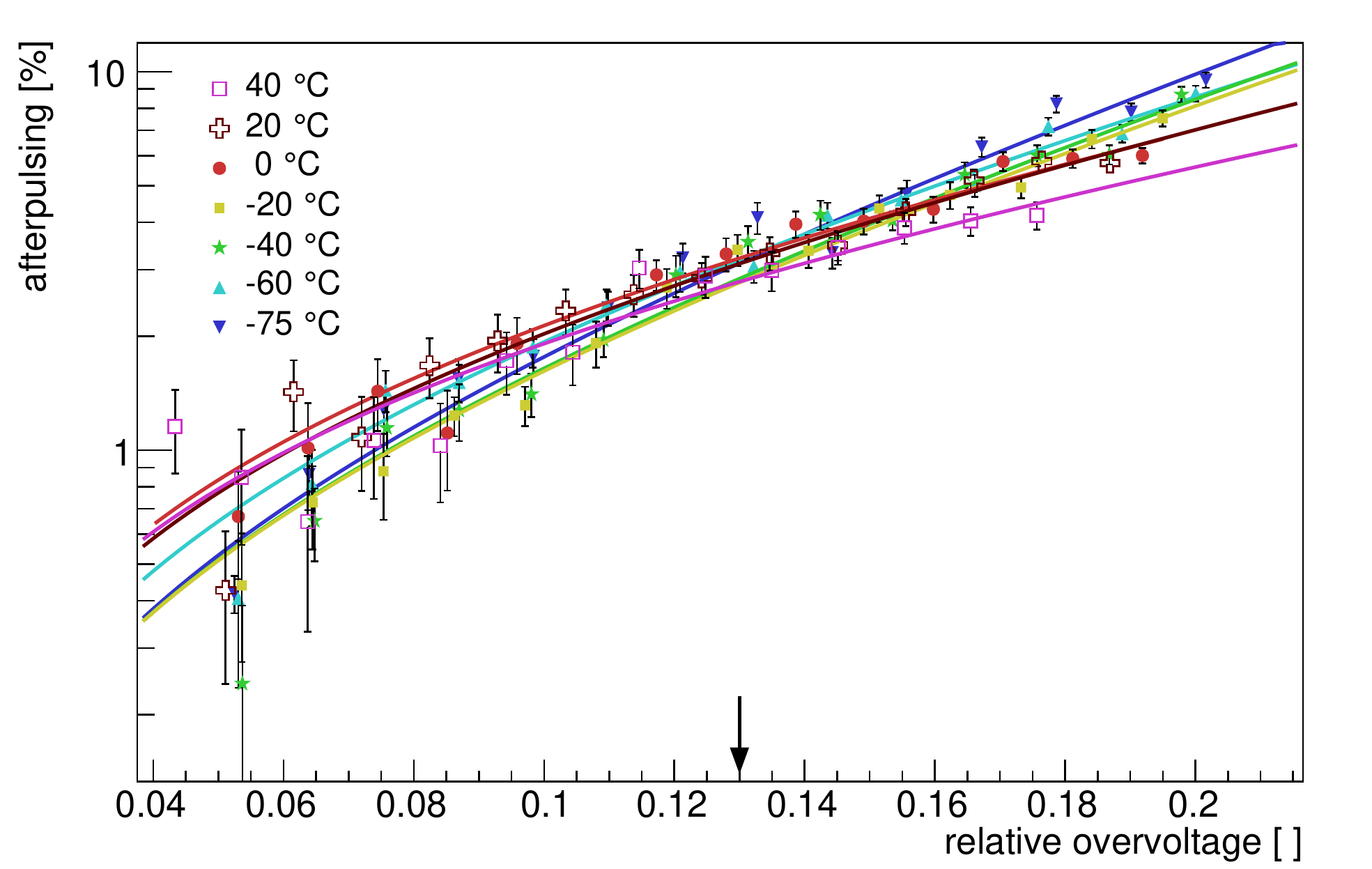}
}
   \subfloat[KETEK PM3325]{
  \includegraphics*[width=\columnwidth]{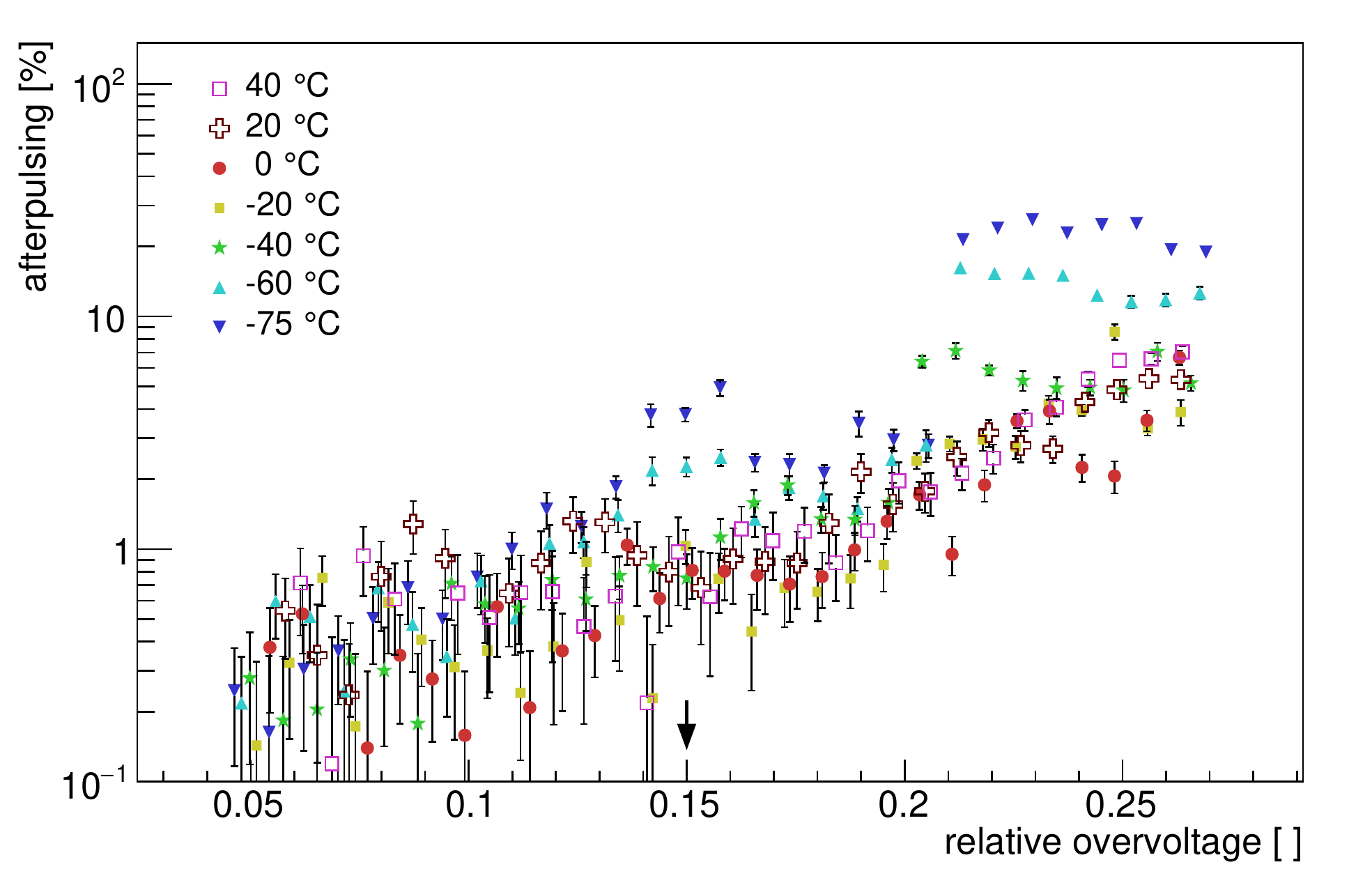}
}
\caption{Afterpulsing probability of the 
two devices.
 The black arrow marks
the relative overvoltage at which both devices yield a 90\% breakdown
probability for 400\,nm photons.
}

  \label{fig:AP}
\end{figure*}

\section{Afterpulsing and Delayed Optical Crosstalk}

If the prompt OC in the Hamamatsu device is indeed dominated by misidentified
delayed OC, a reduction of the minority carrier lifetimes in the bulk with a low
resistivity bulk or a better shielding of the active volume from carriers
diffusing out of the bulk with a potential barrier might be a viable way to
reduce OC further, unless those measures are already implemented. We illustrate
the potential room for improvement by discussing the delayed OC and afterpulsing
characteristics of the two tested SiPMs.

\begin{figure*}[!tb]
  \centering
   \subfloat[Hamamatsu
LVR2]{
  \includegraphics*[width=\columnwidth]{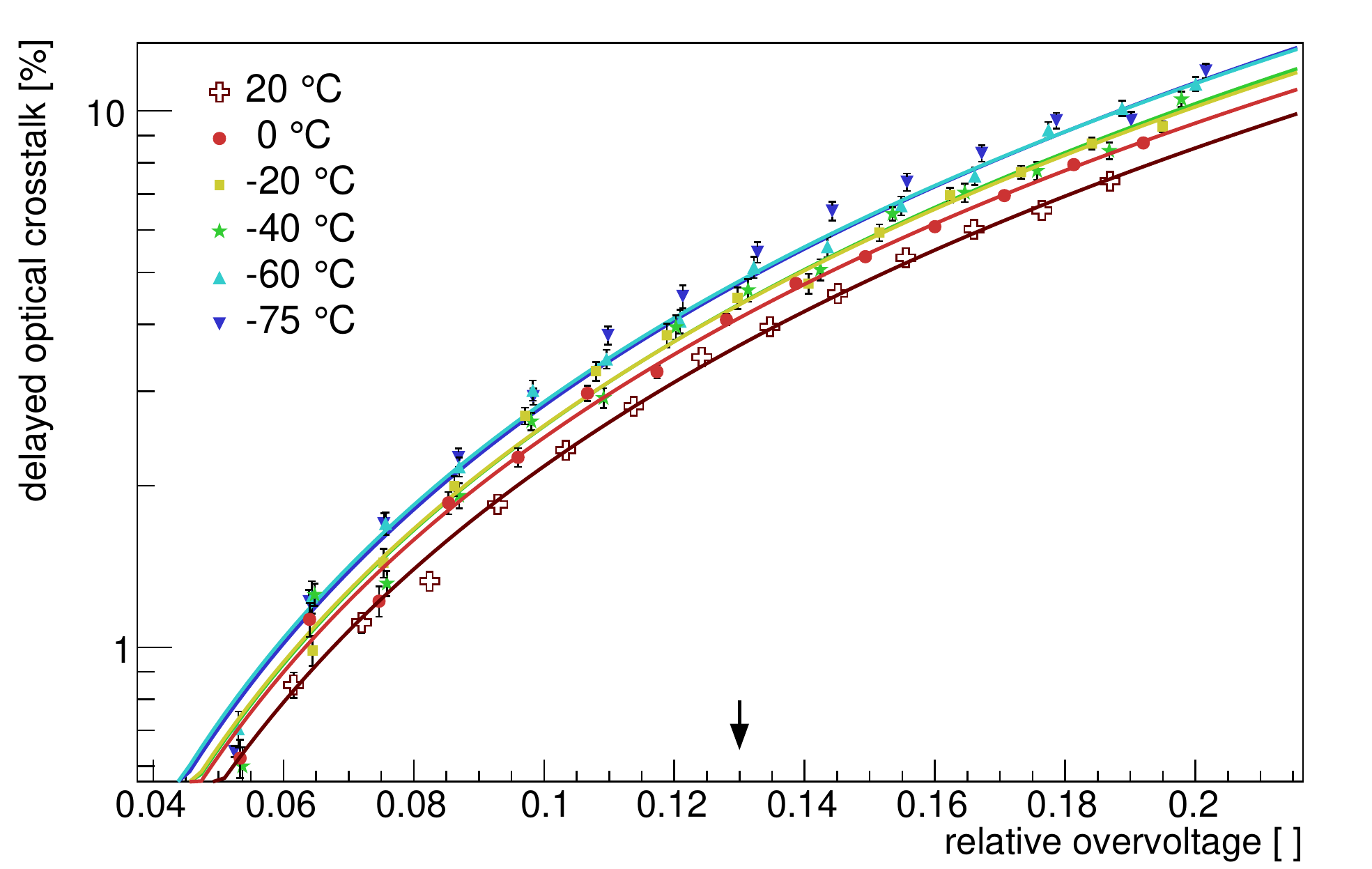}
}
   \subfloat[KETEK PM3325]{
  \includegraphics*[width=\columnwidth]{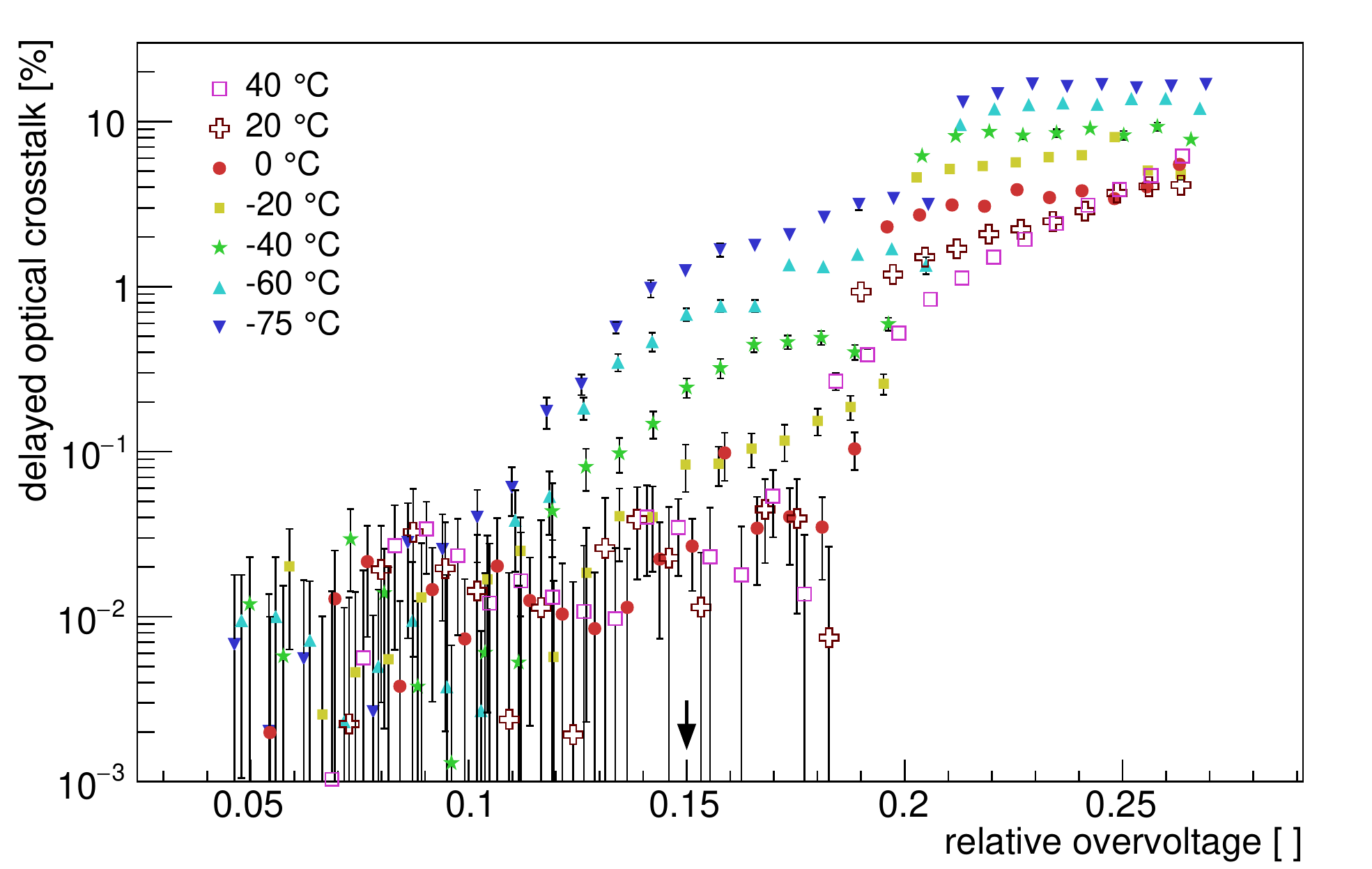}
}
\caption{Delayed optical crosstalk of the 
two devices. 
 The black arrow marks
the relative overvoltage at which both devices yield a 90\% breakdown
probability for 400\,nm photons.
}
  \label{fig:DOC}
\end{figure*}

Both quantities are extracted by recording time difference between SiPM pulses
as explained in \cite{Otte2016a}.  Afterpulsing events become dominant a few ten
nanoseconds after a breakdown when the corresponding cell is recharged to 50\%
or more of its full capacity. Delayed OC signals dominate at shorter time
differences. For the Hamamatsu LVR2 device the subjective division between the two
contributions is made at 20\,ns, and for the KETEK device at 10\,ns. We note
that our choice of separating the two contributions in the described way results
in a contamination of each measurement with events of the opposite type.
That contamination is
acceptable for our purposes.  Figure \ref{fig:AP} shows the afterpulsing and
Figure \ref{fig:DOC} the delayed optical-crosstalk probabilities of both
devices.

The KETEK device has an afterpulsing probability of less than 1\%, whereas the
afterpulsing of the Hamamatsu device is two to three times larger, when
compared at their respective bias, which yields a  90\% breakdown probability
for 400\,nm photons (marked by the arrow in the figures).  
The uncertainties in the different fits of the Hamamatsu
afterpulsing data do not allow us to claim a temperature dependence. The
afterpulsing of the KETEK SiPM shows irregular behavior for relative
overvoltages above 0.2 for the two lowest temperatures. We attribute that
behavior to delayed optical crosstalk leaking into the afterpulsing measurement due to our
choice of discriminating between the two by means of applying a simple cut in time.

At the same 90\% breakdown-probability yielding bias, the delayed OC changes from 0.01\% at $40^\circ$C to
1\% at $-75^\circ$C for the KETEK SiPM.  The temperature dependence is not that
strong in the Hamamatsu SiPM, where the delayed OC is 3.5\% at $20^\circ$C and
increases by a factor of 1.3 to 4.5\% at $-75^\circ$C. We discard the delayed OC
measurement at $40^\circ$C for the same reason we discarded the prompt OC measurement
at the same temperature. 

Below relative overvoltages of 0.15, afterpulsing and delayed OC of the KETEK
device are so low that the measurement is affected by systematic effects.  Only
at higher overvoltages is it possible to resolve the expected temperature
dependence of the delayed optical crosstalk. The dependence is due to an
increase of the carrier life times in the bulk with decreasing temperatures. 

Comparing the prompt and delayed OC performance of both devices has us speculate
about possible future improvements of both technologies. The about ten times
lower delayed OC of the KETEK device is an indication that it should be in
principle possible to lower the delayed OC in the Hamamatsu technology further.
If a lower delayed OC is achieved in the Hamamatsu technology and our assertion
that the prompt OC in the present Hamamatsu device is due to misidentified
delayed OC events, the \emph{effectively measured} prompt OC should go down as well.
On the other hand, it can be expected that future KETEK developments with
trenches will be able to achieve a similar if not better prompt OC performance
than observed in the Hamamatsu SiPM.

\section{Discussion}

In this work we characterized one prototype SiPM from Hamamatsu and the PM3325
WB SiPM from KETEK. Both SiPMs have dramatically improved characteristics when
compared to previous devices. The PDE of both devices peaks between 40\% and
50\% and nuisance parameters are significantly reduced. In particular impressive
is the 1.5\% prompt optical crosstalk of the Hamamatsu device, which is four
times lower than in the Hamamatsu LCT5 device \cite{Otte2016a}. Equally
impressive are the low afterpulsing and delayed optical crosstalk of the KETEK
device, which are both less than 1\%.  A device that combines the excellent
features of both SiPMs would result in another significant improvement in the
SiPM technology.

Analysis methods that probe the microphysics of SiPMs help to understand how
SiPMs work and ultimately provide input in the design of future SiPM
developments. For that purpose we discussed how the vertical
structure of the high-field region is mapped with bias dependent breakdown probability
measurements and how such a mapping can be utilized to learn about the origin of
charge carriers relative to the avalanche structure. Using the method we could
show that the prompt OC producing photons in the Hamamatsu SiPM must be absorbed below
the avalanche structure contrary to the LCT5 device where the majority of OC
photons enter the avalanche region from the surface side.
In the KETEK device, the optical-crosstalk photons \emph{illuminate} the
avalanche region of a neighboring cell from the side.  This information will
help to further improve the prompt OC performance in future devices.  We are not
aware of another experimental method that provides the same information. 

The $\mathcal{O}$-method could also be used to identify the spatial origin of charge
carriers produced by delayed optical crosstalk, afterpulsing, and dark counts
relative to the avalanche region.  However, two requirements need to be
fulfilled first. A valid model has to exist that properly describes the bias
dependence of the characteristic of interest and includes the breakdown
probability. And the measurement cannot be contaminated, like, for example, our
delayed optical crosstalk measurement, which also includes some afterpulsing
events. Unless, of course, the model takes these contaminations into account
too.

The empirical mapping of the $\mathcal{O}$ values obtained in PDE measurements
to the photon absorption length allowed us to determine how far below the
surface the avalanche region is located. However, because we have no access to
the structure of the studied devices, we cannot verify the absolute accuracy of
the mapping and the dependence of $\mathcal{O}$ on the size of the avalanche
region.  To verify that assumption and for a more precise probing of the
high-field structure, dedicated test structures are needed for calibration. The
main parameters to vary in these structures are the size of the region and its
location below the surface.  

Analytical modeling that links $\mathcal{O}$ to the
microphysics of the breakdown, like the ionization coefficients and the
electron/hole breakdown initiation ratio, would further improve the
understanding of SiPMs and expand the usability of the method. We hope that this
paper inspires future work in that direction.

\section*{Acknowledgment}
We are grateful to Hamamatsu and KETEK, who have provided us with samples of their latest developments. This research was in part supported by the National Science Foundation under grant no.\ PHYS-1505228. 

\bibliography{newref}

\end{document}